Localized Phonon Densities of States at Grain Boundaries in Silicon

Peter Rez[1]
Tara Boland[2]
Christian Elsässer[3]
Arunima Singh[1]

1. Department of Physics
   Arizona State University
   Tempe
    AZ 85287-1504
2.  School For Engineering of Matter Transport and Energy
   Arizona State University
   Tempe
    AZ 85287-6106
3. Fraunhofer Institute for Mechanics of Materials IWM
   Wöhlerstraße 11
   79108 Freiburg
   Germany

Abstract

Since it is now possible to record vibrational spectra at nanometer scales in the electron microscope it is of interest to explore whether defects such as dislocations or grain boundaries will result in measurable changes of the spectra. Phonon densities of states were calculated for a set of high angle grain boundaries in silicon. Since these boundaries are modeled by supercells with up to 160 atoms, the density of states was calculated by taking the Fourier transform of the velocity-velocity autocorrelation function from molecular dynamics simulations based on new supercells doubled in each direction. In select cases the results were checked on the original supercells with fewer atoms by comparison with the densities of states obtained by diagonalizing the dynamical matrix calculated using density functional theory. Near the core of the grain boundary the height of the optic phonon peak in the density of states at 60 meV was suppressed relative to features due to acoustic phonons that are largely unchanged relative to their bulk values. This can be attributed to the variation in the strength of bonds in grain boundary core regions where there is a range of bond lengths. It also means that changes in the density of states intrinsic to grain boundaries are unlikely to affect thermal conductivity at ambient temperatures, which are most likely dominated by the scattering of acoustic phonons.

Introduction

Recent developments in monochromators and electron energy loss spectrometers for transmission electron microscopy have made it possible to record vibrational spectra, potentially at nanometer resolution, in the electron microscope[1]. Atomic resolution images at energies corresponding to vibrational modes or phonons have been demonstrated in BN[2] and Si[3], by selecting phonon wave vectors corresponding to localized or impact scattering. This raises the intriguing possibility of experimentally measuring localized phonon modes at extended structural defects in crystals and using measured local densities of states to investigate how defects may change important macroscopic thermal transport properties such as thermal conductivity. Localized modes associated with a single silicon atom in a graphene sheet have been observed by Hage et al[4], and a change in phonon density of states corresponding to a stacking fault in 3C SiC was reported by Yan et al[5].

Localized modes associated with a point defect, namely an isotope atom with a different mass, are to be expected, though the spectacular result of Yan et al [5] showing an observable change at a stacking fault in 3C SiC could be as well a consequence of the stacking fault being a 2 layer region of 2H SiC, a different polytype. Although grain boundaries or dislocations should give rise to differences in the phonon density of states (DOS), it is still not certain whether they would be detectable unequivocally. The energy resolution of the monochromator spectrometer combination in the electron microscope is about 5 meV or 40 cm$^{-1}$; this is a significant advance on what as been achieved in the past, but still very much inferior to what is possible with infra-red absorption or Raman spectroscopy on bulk specimens.

Ziebarth et al[6] and Stoffers et al[7] performed first-principles calculations, using density functional theory (DFT), to determine the atomic positions in supercells representing various grain boundaries in Silicon. For a $\Sigma$ 9 boundary Stoffers et al[7] also compared the DFT results with high resolution scanning transmission electron microcopy images. Since Venkatraman et al[3] have shown that it is possible to measure the phonon scattering at the atomic scale in Silicon using an on-axis detector, any local changes at grain boundaries may be experimentally detectable.

We used the atomic coordinates of the structurally relaxed grain-boundary models of Ziebarth et al[6] and Stoffers et al[7] to calculate the phonon DOS in two ways: by enumerating the modes obtained by diagonalizing the dynamical matrix, calculated directly by DFT; by taking the Fourier transform of the velocity-velocity correlation function of the atoms in an extended molecular dynamics (MD) simulation. In calculating the DOS we made a distinction between those atoms at the grain boundary core and those that were in regions that can be considered as similar to the bulk Si crystal. In all cases except the $\Sigma$ 5 (130) boundary the peak in the DOS at 60 meV was depressed for atoms in the core region. We attribute the anomalous result for the $\Sigma$ 5 (130) boundary to the small size of the supercell which means that the environment of all the atoms is more like that of the grain boundary core than that of the bulk crystal.

Theory

Phonons are the quantized vibrations of the atoms in the crystal lattice; they are uniquely determined by their wave vector in the 1$^{st}$ Brillouin Zone and their polarization vector. Since the grain boundary systems that we are studying are represented by large supercells, the Brillouin Zone is relatively small, and as a first approximation the density of states needs only to be evaluated by enumerating the frequencies at the Γ point (or typically only a small number of k points) [8]. The frequencies are obtained by diagonalizing the dynamical matrix $D$

$$D\begin{pmatrix} k \\ b & b' \\ \alpha & \alpha' \end{pmatrix} = \frac{e^{-i\mathbf{k}.\mathbf{R}_{b'}}}{\sqrt{M_{b'}}} \phi \begin{pmatrix} b & b' \\ \alpha & \alpha' \end{pmatrix} \frac{e^{i\mathbf{k}.\mathbf{R}_b}}{\sqrt{M_b}} \qquad 1$$

where $b$ labels atoms of mass $M_b$ in the simulation cell with periodic boundary conditions, $\alpha$ labels the polarization and $\Phi$ is the force-constant matrix, and $k$ is the wave vector. The number of modes is $3N$ where $N$ is the number of atoms in the unit cell. The original supercells that represent our grain boundaries contain up to 160 atoms, thus the dynamical matrix can be of size up to 480 x 480. All DFT calculations were done using the projector-augmented-wave method as implemented in the plane-wave code VASP [9-11] within the generalized gradient approximation of Perdew, Burke and Ernzerhof for exchange-correlation.[12] A plane-wave energy cutoff of 400 eV was used. A $k$-point sampling mesh with at least 10 points per Å$^{-1}$ was used for the supercells together with a 0.01 eV smearing width of the Gaussian smearing scheme. The grain-boundary (GB) models of Ziebarth et al[6] and Stoffers et al[7] were re-relaxed using the aforementioned parameters. These settings were sufficient to converge all five GB cells to a total force per atom of less than 0.05 eV/ Å or better and an energy tolerance of $10^{-5}$ eV. A compilation of the lattice parameters of the cells and the numbers of atoms is given in Table 1. Phonon calculations were performed using the direct finite-difference method[13,14] implemented within the Phonopy package [15]. To obtain the projected phonon density of states, a 1x1x1 supercell approach was used for all GB cells with a 10x10x10 k-point sampling mesh density. Finite atomic displacements of 0.01 Å were applied to calculate the force-constant matrix. The atomic displacements of the phonon modes, obtained from the eigenvectors of the modes at the Gamma point, were simulated using the v_sim code (https://gitlab.com/l_sim/v_sim)

Table 1   Compilation of numbers of atoms and unit cell parameters of the supercells for the five investigated grain boundaries. The "large" MD supercells contain $2^3=8$ times as many atoms and have unit-cell parameters twice as long.

|  | Σ3 (112) Mirror-symmetric | Σ3 (112) Non-symmetric | Σ5 (130) | Σ5 (120) | Σ9 (221) |
|---|---|---|---|---|---|
| # of atoms in DFT supercell | 144 | 136 | 40 | 160 | 68 |
| # of atoms in MD supercell | 1152 | 1088 | 320 | 1280 | 544 |
| a (Å) | 9.4790 | 9.4336 | 5.3335 | 12.3261 | 11.6531 |
| b (Å) | 40.5757 | 38.7540 | 17.6788 | 24.5705 | 31.0746 |
| c (Å) | 7.6715 | 7.6715 | 8.7036 | 10.8160 | 3.8576 |

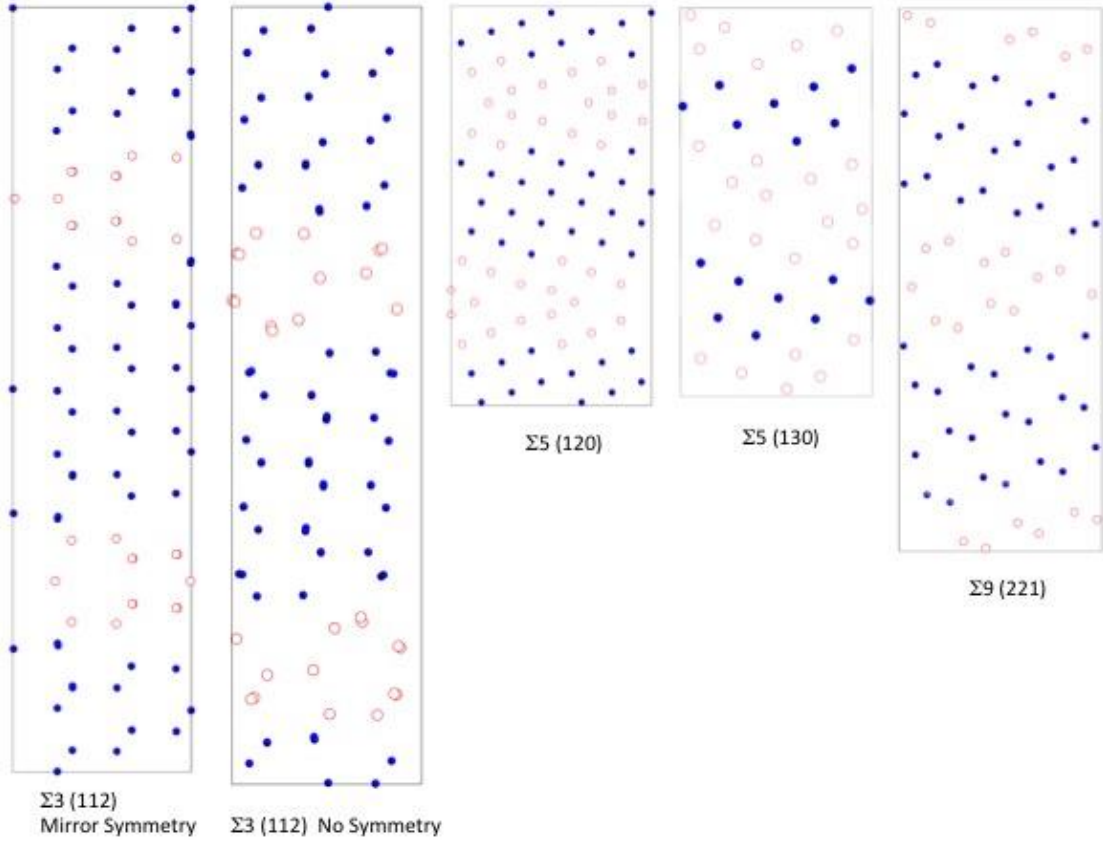

Figure 1 The original DFT supercells for the various grain boundaries. The atoms at the grain boundary core are shown as open red circles and the "bulk" atoms are shown as filled blue circles.

An alternative method to calculate the phonon DOS is to take the Fourier transform of the velocity-velocity correlation function from MD trajectories of the atoms[16,17,18]. The instantaneous velocities are determined by the differences in the atom positions in the time evolution of the MD trajectories. In practice it was found to be beneficial to break the trajectory into $L$ time slices of $M$ steps and average over the resulting spectra. The phonon density of states $I(\omega_k)$ is

$$I(\omega_k) = \frac{1}{L}\sum_{l=0}^{l=L-1}\sum_{j=2+lM}^{j=(l+1)M} exp(i\omega_k t_j) \sum_{i=1}^{i=p}\sum_{n=1}^{n=N} v_{in}(t_j)v_{in}(t_2) exp\left(-\frac{\sigma^2 t_j^2}{4}\right)$$

where $v_{ik}(t_j)$ is proportional to the velocity at time $t_j$

$$v_{ik}(t_j) = u_{ik}(t_j) - u_{ik}(t_{j-1})$$

and $M$ is the number of time steps in each time slice $l$, $t_j$ is the $j^{th}$ time step, $N$ the number of atoms, $\omega_k$ is a frequency, $u_i$ is the coordinate in the $i$'th direction (x,y,z) for atom $k$. For the total density of states it is summed over dimensions x,y and z (p=3). The parameter $\sigma$ is related to the instrumental broadening, $\Delta E$ by

$$\sigma = \frac{\Delta E e}{2\hbar\sqrt{\ln 2}}$$

when $\Delta E$ is in eV.

This approach was proposed by Beeman and Alben [19] and has been used to calculate the thermal conductivity of MgO by de Koker[20], phonon dispersions and lifetimes in carbon nanotubes by Thomas et al, [21] and infra red spectra of large proteins by Mott et al [22].

Scattering wave vectors in electron energy loss spectroscopy (EELS) in the electron microscope are predominately in the plane of the specimen, normal to the incident beam. For densities of states that would be observed by EELS in the electron microscope the displacement direction, $i$, is only summed over the 2 dimensions in the plane of the specimen (p=2). The specimen's surface plane normal for the $\Sigma$ 3 and $\Sigma$9 boundaries in Fig 1 is (110), which is the most frequently used orientation for high-resolution electron microscopy of silicon[23,24]. There have been fewer high-resolution electron microscopy results for the (100) orientation which is the specimen normal for the $\Sigma$ 5 boundaries[25].

Molecular dynamics simulations were performed with the LAMMPS [26] package using the phonon fix command that measures the elastic Green's function by making use of the fluctuation-dissipation theorem implemented by Kong [27] to obtain the phonon spectrum. The positions of atoms in the supercells of grain boundaries determined by Ziebarth et al [6] and Stoffers et al [7] were used as the initial atom coordinates. In practice it was found that their supercells were too small for calculating the phonon DOS. Various multiples of the supercell were tried, and it was found that convergence was achieved with a larger supercell containing the original supercells 2 x 2 x 2 times. The numbers of atoms in these larger supercells are given in the 2$^{nd}$ line of Table 1. These simulations were performed with periodic boundary conditions applied in all directions. The interaction among atoms was described using the Tersoff potential [28]. The Nosé–Hoover style thermostat and barostat [29-32] were employed to maintain the system at 300 K and 1 atmosphere. A time step of 1 fs was used with a total simulation time of 0.05 ns. The atom positions were saved at each time step. These computational settings were sufficient to obtain a stable spectrum. Additional convergence checks were performed for simulation times of 0.1 ns. For calculating the spectra as given by equations 2 and 3, the trajectories were divided into 24 blocks of 2048 time steps and the results were averaged. All spectra were normalized using the constraint that there are 3 vibrational modes per atom.

Calculating phonon density of states by DFT and Phonopy for the supercells shown in Fig 1 is computer resource and time intensive. This was only done for selected cases as a way of checking and validating the results from the analysis of molecular dynamics trajectories by direct comparison with the calculation of the phonon spectrum from the dynamical matrix.

Results and Discussion

To validate the molecular dynamics approach, in particular the Tersoff potential we used, we compared the phonon density of states calculated by both molecular dynamics and by diagonalizing the dynamical matrix using VASP in association with Phonopy.  The result for a perfect crystal is shown as Fig 2.  The most prominent feature is the peak in the DOS at 60 meV. Silicon has an optic mode since there are two atoms in each primitive unit cell.  There is little dispersion with wave vector for optic modes, which means they make a large contribution to the density of states. The peaks and their relative heights are the same, though the peaks in the DOS calculated using molecular dynamics trajectories with the Tersoff potential are shifted to slightly higher energies. This is not unexpected, a similar shift was reported by Rohskopf et al[33]. However the strong similarity between the results of the two different approaches suggests that taking the Fourier transform of the velocity-velocity correlation function of instantaneous velocities from MD trajectories is a valid approach, though it should be remembered that the Tersoff potential is constructed originally for the perfect crystal structure and might be of limited validity for the distorted structures at grain boundaries.

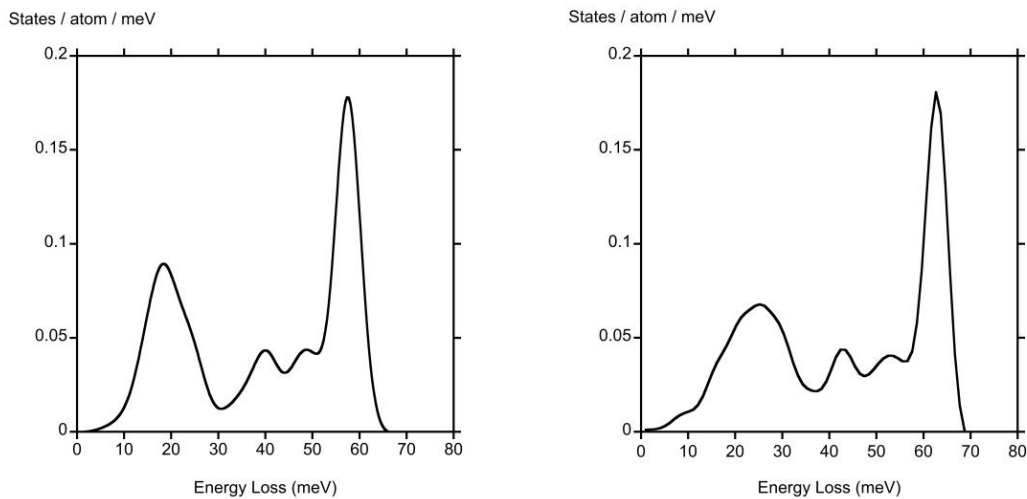

Fig 2  Phonon density of states for perfect Si: (a) calculated with DFT (using VASP and Phonopy) (b) calculated from molecular dynamics trajectories and the Tersoff potential.

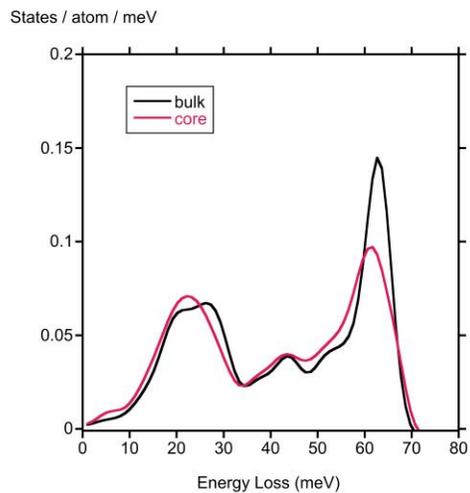

(a) Σ 3 (112) mirror-symmetric GB, MD total DOS

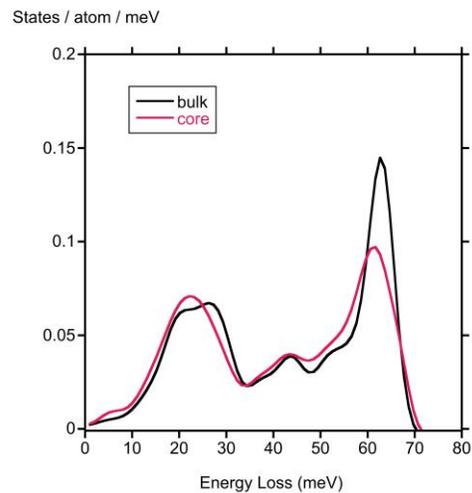

(b) Σ 3 (112) non-symmetric GB, MD total DOS

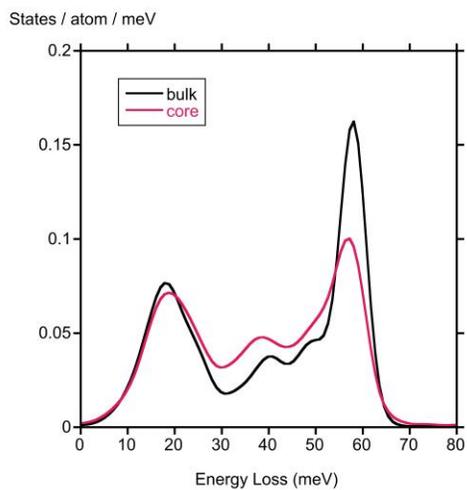

(c) Σ 3 (112) mirror-symmetric GB, DFT total DOS

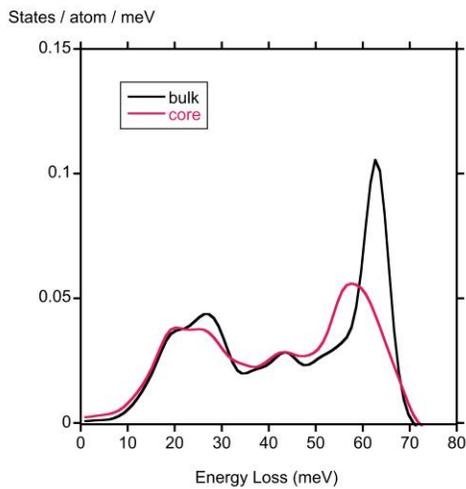
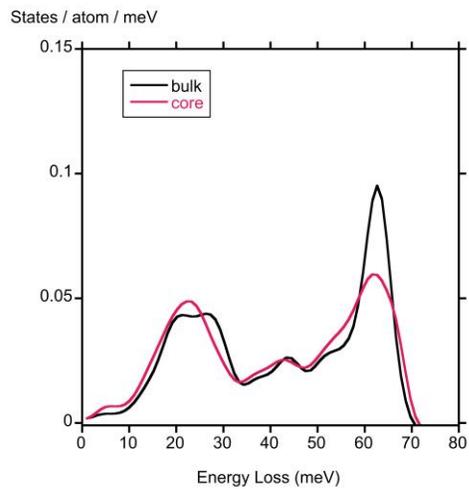

(d) Σ 3 (112) mirror-symmetric GB, MD EELS DOS

(e) Σ 3 (112) non-symmetric GB, MD EELS DOS

Figure 3   Phonon densities of states for Σ 3 (112) grain boundaries. Graphs (a), (b) and (c) display the total DOS, graphs (d) and (e) illustrate the partial DOS corresponding to EELS spectra.

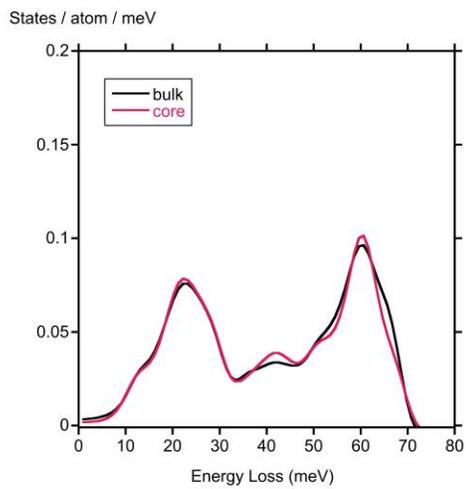

(a) Σ 5 (130) GB, MD total DOS

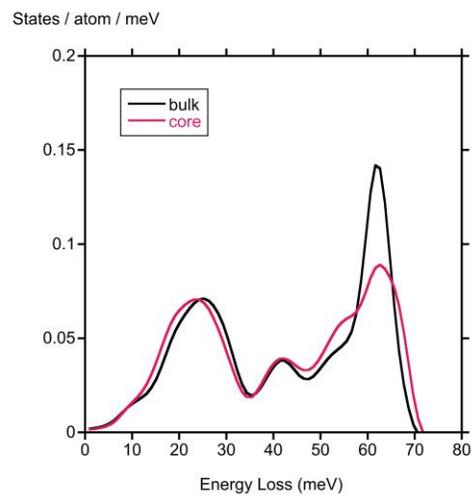

(b) Σ 5 (120) GB, MD total DOS

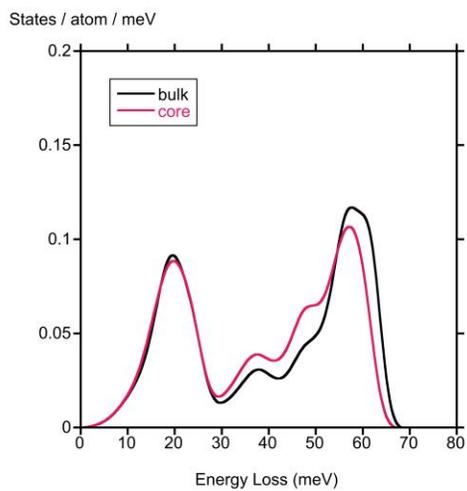

(c) Σ 5 (130) GB, DFT total DOS

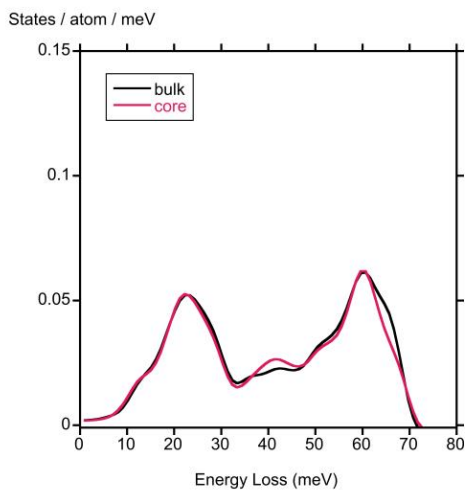
(d) Σ 5 (130) GB, MD EELS DOS

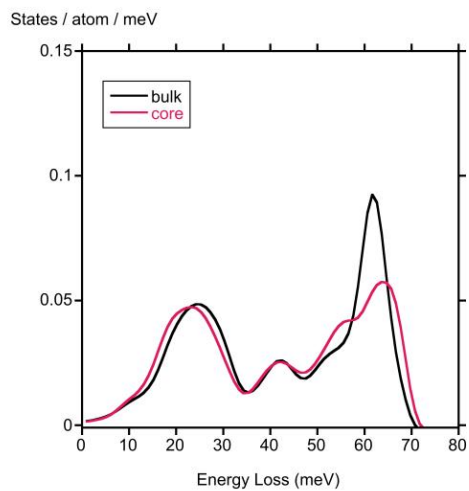
(e) Σ 5 (120) GB, MD EELS DOS

Figure 4    Phonon densities of states for Σ 5 grain boundaries. Graphs (a), (b) and (c) display the total DOS, graphs (d) and (e) illustrate the partial DOS corresponding to EELS spectra.

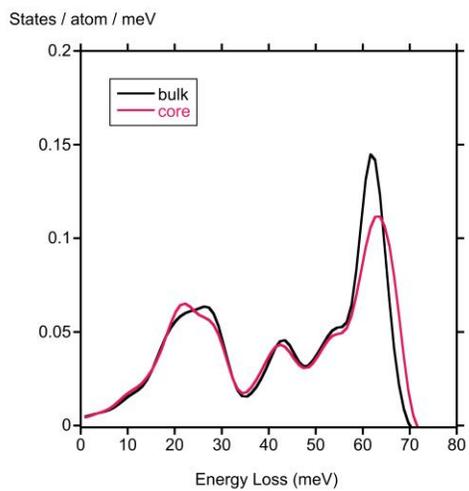

(a) Σ 9 (221) GB, MD total DOS

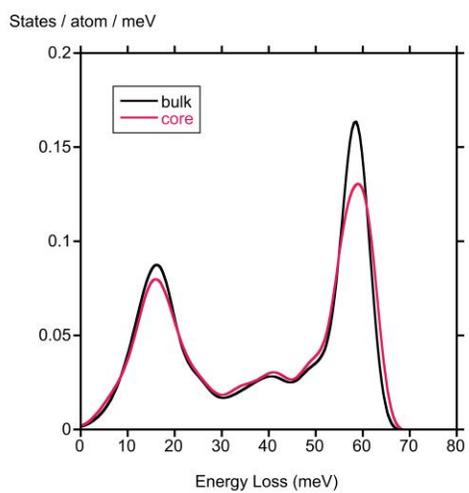

(b) Σ 9 (221) GB, DFT total DOS

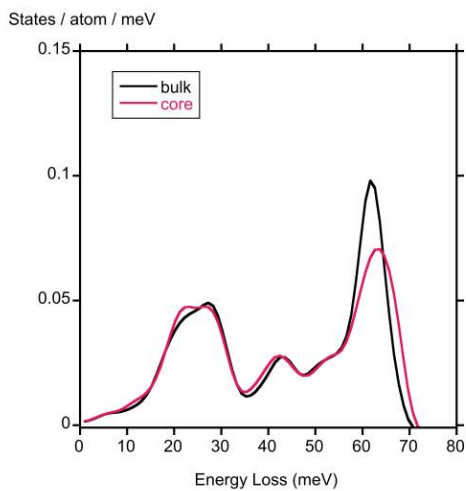

(c) Σ 9 (221) GB, MD EELS DOS

Figure 5  Phonon DOS results for Σ 9 (221) grain boundary.  Graphs (a) and (b) display the total DOS, graph (c) illustrates a partial DOS corresponding to an EELS spectrum.

Figures 3-5 display comparisons between the total phonon density of states and the partial density of states that would be observed by EELS for atoms that are part of a grain boundary (GB *core* atoms) and those that are more distant from the boundary (*bulk* atoms).  The core and bulk atoms are marked as open red and filled blue circles respectively in Figure 1.  In all the graphs the density of states is broadened by a Gaussian of 5 meV half width to represent the instrumental energy resolution that can now be achieved in a transmission electron microscope.  Given that it is possible to record spectra from regions as small as 0.5 nm, it should be possible to acquire distinct spectra representative of the core and bulk regions.

In all cases the most significant and observable change is the reduction of the height of the optical-phonon peak in the DOS at 60 meV.  The magnitude of the reduction in peak height is different for the different boundaries.  It is greatest for the Σ 3 (112) mirror-symmetric boundary, and hardly noticeable for the Σ5 (130) boundary. The effect for the boundary with the lowest tilt angle (or highest sigma value) considered, the Σ 9 (221) boundary, is somewhere in between that for the Σ 3 (112) mirror-symmetric boundary and the Σ 5 (130) boundary.  It is interesting to note that the projected DOS that would be observed by EELS in the electron microscope is very similar to the total DOS.  The magnitude of the DOS in states/atom/meV is 2/3 of the 3D DOS, as EELS selects a 2D projection.  We performed DFT (VASP & Phonopy) calculations for the Σ 3 (112) mirror-symmetric boundary, the Σ 5 (130) boundary and the Σ 9 (221) boundary.  These were intentionally chosen to span the range of changes in peak heights obtained by the MD calculations. Although the exact values of the height reduction in the 60 meV peak differ between the DFT and MD calculations, most noticeably in the case of the Σ5 (130) boundary, the same trend is apparent.

The magnitude of the peak at 60 meV is controlled by the angle of the boundary and the Coincidence Site Lattice misorientation, as measured by the parameter Σ. The nature of the boundary plane can have a significant effect as shown by the difference between the peak height for the Σ 5 (130) and (120) boundaries and between the Σ 3 (112) boundaries with and without mirror symmetry. However for the Σ 5 (130) boundary the separation between the boundaries in the supercell was so small that most of the atoms are part of the boundary cores, as can be seen in Fig 1. Interatomic distances and angles vary considerably in the core regions of the grain boundary, and this is reflected in the interatomic potential and the magnitude of the interatomic forces. This is very evident from the map of displacements for the mirror symmetric Σ 3 (112) and the Σ 9 (221) boundaries shown as Figs 6a and b, where the displacements are much smaller for atoms near the boundary core.

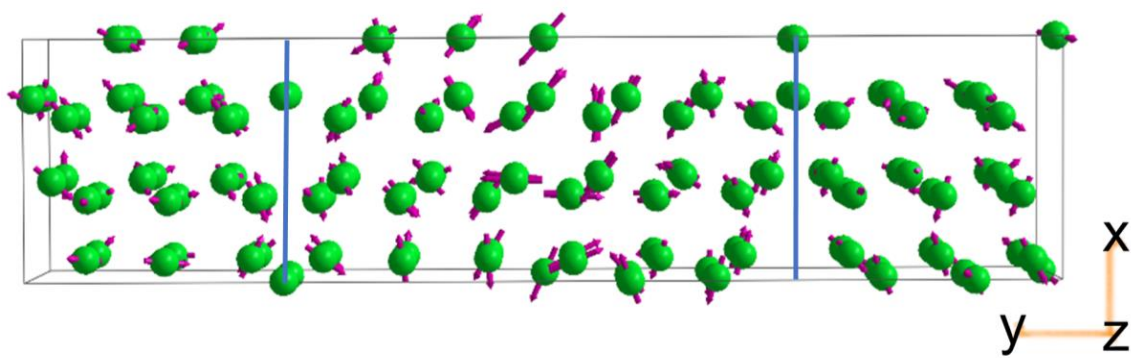

(a) Displacements for the 59 meV mode for the mirror symmetric Σ 3 (112) boundary

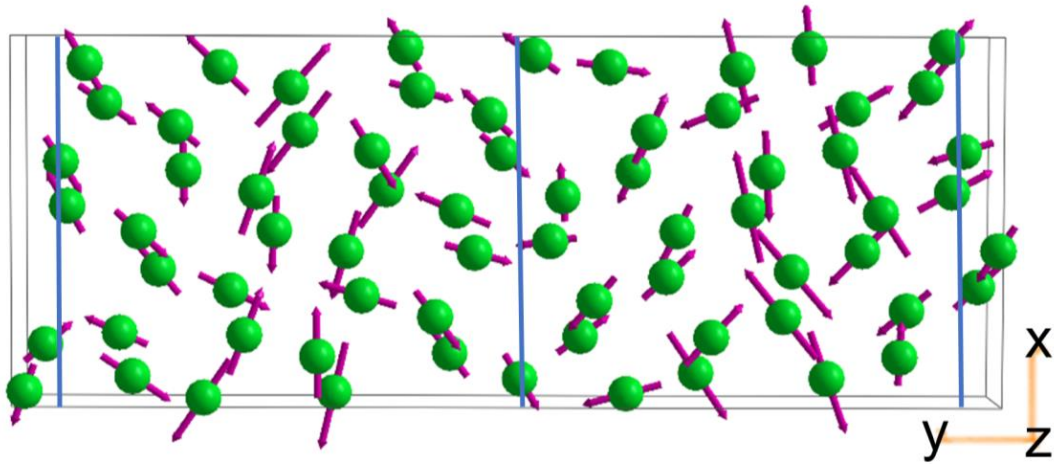

(b) Displacements for the 58 meV mode for the Σ 9 (221) boundary

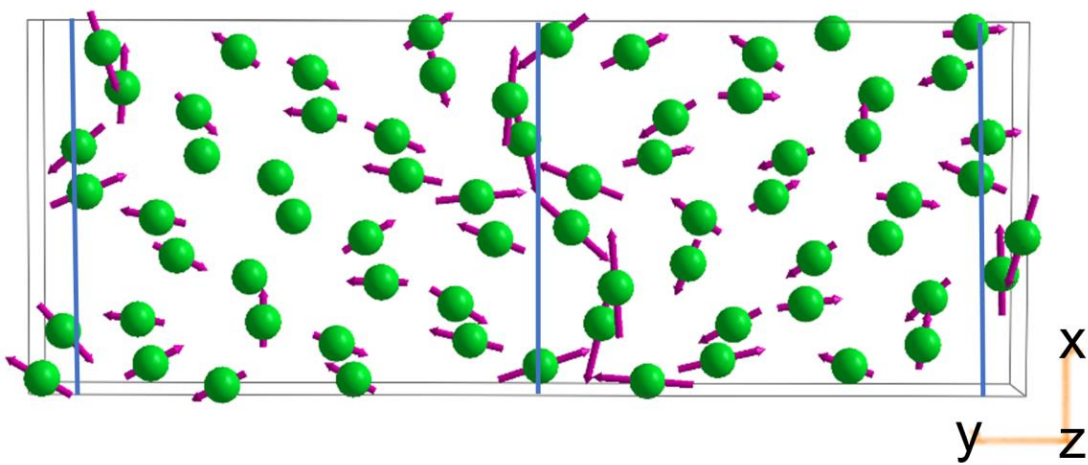

(c)     Displacements for the 63 meV mode for the Σ 9 (221) boundary

Figure 6 Atomic displacements for modes predominately affecting bulk like atoms for (a) the Σ 3 (112) boundary and (b) the Σ 9 (221) boundary, (c) is a mode localized on boundary core atoms of the Σ 9 (221) boundary.  The blue lines mark the locations of the boundary planes in the supercells.

It is even more striking in the animations shown as movies S1 and S2 in Supplementary Information.  Since the magnitude of the DOS is related to the size of the atomic displacements as can be seen from equations 2 and 3 it is not surprising that the 60 meV optic phonon peak is reduced in both the total DOS and the projected DOS observed by EELS.  There are relatively few modes localized to atoms at the core of the boundaries. There is a single mode at 63 meV for the Σ 9 (221) boundary with the displacements shown as Fig 6c. The animation is shown as movie S3 in the Supplementary Information. However it is unlikely that this single mode would be separately detected when the instrumental resolution is 5 meV.

Conclusions

Phonon densities of states in silicon calculated from velocity-velocity correlation functions of molecular dynamics trajectories (using a Tersoff potential) were validated by comparison with those calculated directly from the dynamical matrix by density functional theory (using VASP and Phonopy).  This agreement is not only true for perfect crystals but also for extended structural defects, namely grain boundaries, where there might be concerns about the applicability of simple force fields.  The phonon density of states due to optic phonons at 60 meV for Σ3 to Σ9 high-angle tilt boundaries is reduced compared to the optic phonon density of states in regions of bulk crystal structure.  This change should be detectable by high resolution EELS in electron microscopes in a specimen where the boundary plane is perpendicular to the specimen surface, and the electron beam runs parallel to the boundary core. The similarity between the densities of states from acoustic phonons at the boundary core and in bulk regions indicates that local changes of densities of states at the grain boundaries cores should have minimal effect on the thermal conductivity of a polycrystalline microstructure of Si at ambient temperature. Any changes in thermal conductivity will be a consequence of phonon scattering at the grain boundary leading to a reduction in phonon mean free paths.

Acknowledgements

T. B. and A. S. are supported in part by the Arizona State University start-up funds. A. S. is also funded by the NSF DMR grant #1906030 and as part of ULTRA, an Energy Frontier Research Center funded by the U.S. Department of Energy (DOE), Office of Science, Basic Energy Sciences (BES), under Award # DE-SC0021230. This research used computational resources from the NSF-XSEDE under Award No. DMR150006, HPC resources from the Research Computing at Arizona State University, and computational resources from the National Energy Research Scientific Computing Center, a DOE Office of Science User Facility supported by the Office of Science of the U.S. Department of Energy under Contract No. DE-AC02-05CH11231.